\documentclass{article}

\usepackage[final,main]{neurips_2025}

\usepackage[utf8]{inputenc}
\usepackage[T1]{fontenc}
\usepackage[hidelinks]{hyperref}
\usepackage{url}
\usepackage{booktabs}
\usepackage{amsfonts}
\usepackage{amsmath}
\usepackage{amssymb}
\usepackage{nicefrac}
\usepackage{microtype}
\usepackage{xcolor}
\usepackage{graphicx}
\usepackage{algorithm}
\usepackage{algpseudocode}
\usepackage{float}
\usepackage{subcaption}

\title{GraphGP: Scalable Gaussian Processes with \\Vecchia's Approximation}

\author{%
  Benjamin Dodge \quad Philipp Frank \quad Susan E. Clark \\
  Kavli Institute for Particle Astrophysics and Cosmology \\
  Stanford University Department of Physics \\
  Correspondence: \texttt{bendodge@stanford.edu}
}

\begin{document}

\maketitle

\begin{abstract}
  Gaussian processes are a powerful tool for modeling continuous fields, but their naive $\mathcal{O}(N^3)$ computational cost and $\mathcal{O}(N^2)$ memory requirement often limit their practical use. Vecchia's approximation is a sparse precision matrix approximation for stationary, decaying kernels that conditions each point only on its $k$ nearest neighbors. We present GraphGP,\footnote{https://github.com/stanford-ism/graphgp} a GPU algorithm for Vecchia's approximation that scales to nearly a billion parameters with linear time and memory requirements, handling arbitrary point distributions over a large dynamic range. Our key contributions are (1) a bit-reversed $k$-d tree ordering that allows efficient neighbor searches while also maximizing batch parallelism, and (2) a differentiable CUDA implementation, which is substantially faster and more memory efficient than our pure JAX baseline. GraphGP provides the building blocks for inference, including forward generation, inverse application, log-determinant, and kernel parameter derivatives.
\end{abstract}

\section{Introduction}
\label{sec:intro}

Gaussian processes (GPs) offer a principled framework for spatial modeling and uncertainty quantification, but scaling them to millions or billions of points remains a challenge. Inducing-point methods reduce cost to $\mathcal{O}(NM^2)$ through $M \ll N$ inducing variables but struggle when $M$ must be large \citep{titsias2009,wilson2015}. Fast Fourier transforms run in $\mathcal{O}(N \log N)$ time on regular grids with stationary covariance kernels, but are difficult to apply to arbitrary point distributions with large dynamic range. Vecchia's approximation~\citep{vecchia1988,katzfuss2020} achieves $\mathcal{O}(Nk^3)$ scaling by conditioning each new value on only $k\ll N$ previously generated values---a good approximation for certain stationary kernels, but difficult to execute in practice for large $N$. Iterative Charted Refinement (ICR)~\citep{edenhofer2022} uses a similar approximation on a hierarchical grid, enabling it to scale to hundreds of millions of parameters on a single GPU.

GraphGP combines the flexibility of Vecchia's approximation on arbitrary points with the scalability that ICR achieves on a grid. Our target application is three-dimensional mapping of the interstellar medium, where ICR-based GP priors have been used with great success~\citep{leike2022,edenhofer2024,zandinejad2026,mccallum2026}. GraphGP will enable a new generation of these maps with inhomogeneous resolution and lower memory requirements, among other potential applications in Bayesian inference of fields across the physical sciences. The main challenges we address in this paper are how to quickly determine appropriate conditioning relationships and parallelize the resulting computation, both of which are accomplished using a new generation order (Section \ref{sec:ordering}). We also identify several CUDA optimizations that dramatically reduce time and memory requirements compared to our JAX baseline (Section~\ref{sec:implementation}). Finally, we conclude and compare with prior work (Section~\ref{sec:discussion}).

\section{Algorithm}
\label{sec:algorithm}

\subsection{Vecchia's Approximation}
\label{sec:vecchia}

Given $N$ points $\mathbf{x}_1, \ldots, \mathbf{x}_N \in \mathbb{R}^d$ and a stationary covariance $\kappa(r)$, we seek to sample $\mathbf{v}\sim\mathcal{N}(0, K)$ where $K=\kappa(\|\mathbf{x}-\mathbf{x}^\top\|)$. Vecchia's approximation~\citep{vecchia1988} factors the joint density as

\vspace{-0.3cm}

\begin{equation}
  p(\mathbf{v}) \approx \prod_{i=1}^{N} p(v_i \mid \mathbf{v}_{c(i)})
  \label{eq:vecchia}
\end{equation}
where $c(i) \subset \{1, \ldots, i{-}1\}$ is a set of preceding conditioning points. Each factor is Gaussian and can be computed with the Gaussian conditioning formulas, where we have written $c=c(i)$ for clarity.

\vspace{-0.3cm}

\begin{equation}
  v_i \mid \mathbf{v}_{c} \sim \mathcal{G}(K_{ic} K_{cc}^{-1}\mathbf{v}_{c}, \: K_{ii}-K_{ic}K_{cc}^{-1}K_{ci})
  \label{eq:conditional}
\end{equation}

Samples are generated sequentially by drawing each $v_i$ from this distribution. If $c(i)$ contains all points up to $i-1$, the approximation becomes exact and equivalent to using the dense lower-triangular Cholesky factorization $K=LL^T$. Instead, we only condition on $k\ll N$ preceding points, dramatically reducing the size of the conditioning matrices so that sampling takes only $\mathcal{O}(Nk^3)$ time and $\mathcal{O}(Nk^2)$ memory. Often, the $k$ nearest neighbors among preceding points are chosen, since for decreasing $\kappa(r)$ these independently explain the most variance of $v_i$. Note that this is different from truncating the covariance to the range of these neighbors since long-range correlation can propagate via conditionals. The effective covariance is dense while the inverse covariance, or precision, is sparse. See \cite{katzfuss2020} for a more detailed introduction to Vecchia's approximation.

\subsection{Ordering}
\label{sec:ordering}

Vecchia's approximation depends not only on the number of neighbors $k$, but also the order of generation since neighbors are chosen among preceding points. This order has significant implications for the scalability of the resulting algorithm. In this section, we propose an order that (1) allows fast querying of \textit{preceding} neighbors and (2) enables many values to be generated in parallel.

\paragraph{Tree ordering.} Nearest neighbor algorithms often rely on a $k$-d tree data structure that recursively partitions points in space. When querying neighbors, branches of the tree can be pruned if their points are guaranteed to be farther than the current candidate neighbors. Unfortunately, this does not allow efficient queries among the small subset of preceding points as most of the tree must be traversed until sufficient points are found. Instead, we observe that in a left-balanced $k$-d tree in binary tree order, every prefix sequence is also a valid left-balanced $k$-d tree. Therefore we can use tree order as the generation order so that preceding points are already in a tree and can be queried efficiently.

\begin{figure}
  \includegraphics[width=\linewidth]{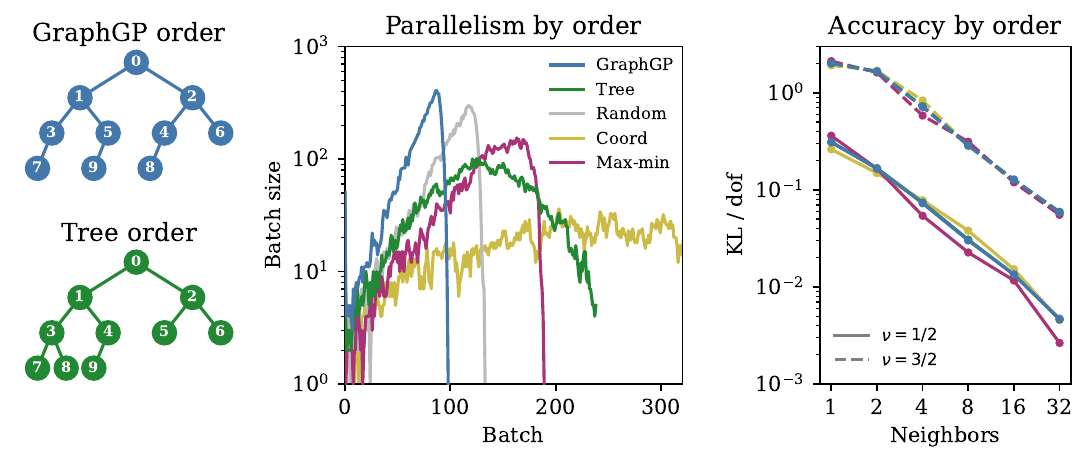}
  \caption{Evaluation of GraphGP order. \textit{Left:} Illustration of our bit-reversed order and the standard binary $k$-d tree order. Consecutive points in our order are far away from each other in space and likely to be conditionally independent. \textit{Center:} Batch sizes for $N=10^4$ points drawn from a unit Gaussian in 3d with $k=16$ and $n_0=50$. We choose this test distribution as it is irregular and offers three orders of magnitude in dynamic range while being small enough for dense matrix operations. We compare the GraphGP order, standard tree order, random order, $x$-coordinate sorted order, and the ``max-min'' order proposed by \cite{guinness2018}. Our order offers the largest parallel batches while simultaneously enabling scalable preceding neighbor searches. \textit{Right:} Accuracy as measured by the Kullback–Leibler divergence from truth for a Matérn covariance with unit variance, cutoff $\rho=10$, and smoothing parameter $\nu$. While the quantitative results will depend on the point distribution and covariance assumed, the impact of order on accuracy is minor compared to the number of neighbors used, justifying our focus on favorable scaling properties.}
  \label{fig:ordering}
  \vspace{-0.5cm}
\end{figure}

\paragraph{Bit-reversed ordering.}  Fast generation requires that we generate many values in parallel, which can only be achieved if large batches of points are conditionally independent (none are each other's neighbors). Unfortunately, this is not the case for the standard $k$-d tree order as nearby points are close together by design and tend to produce long dependency chains. To address this, we propose a modified tree order that inserts all left children before any right children, spreading out nearby points. Specifically, node $i$ has left child $i+2^l$ and right child $i+2^{l+1}$ at level $l$. This is equivalent to a bit-reversal permutation of each level if all levels are full. Importantly, any prefix sequence is still a valid tree in this new order. Figure~\ref{fig:ordering} illustrates the difference and shows that this order results in much larger batches of conditionally independent points. In practice, we find that hundreds of millions of values can be generated in a few hundred batches, easily saturating modern GPUs.


\paragraph{Batch construction.}  Conditionally independent batches are not contiguous and must be identified by analyzing the directed acyclic graph implied by preceding neighbor relationships. Fortunately, the graph is already in topological order by construction (all edges point from earlier to later points). We can therefore compute the depth, or longest path from root, of each node by iterating over the points in order and applying $d_i = 1 + \max_{j \in c(i)} d_j$. Nodes at the same depth can be generated in parallel as they only depend on nodes of lesser depth. We generate values in depth-sorted order, which does not change any conditioning relationships but results in contiguous parallel batches. To avoid small initial batches and increase accuracy, we generate the first $n_0$ values with a dense Cholesky factorization.


\subsection{Implementation}
\label{sec:implementation}

Our baseline implementation of GraphGP is written in JAX \citep{jax2018}, providing automatic differentiation and GPU acceleration with minimal development overhead. Unfortunately, several important optimizations are not possible to achieve with JAX alone, so we also provide a custom CUDA implementation for key operations.

Memory usage in Vecchia's approximation is dominated by the $k \times k$ conditioning matrices required for each point. However, these matrices are only needed for a short time and are often quite small for our target applications ($k \lesssim 64$). Therefore, we can compute them on-the-fly in GPU registers, avoiding the need to allocate global memory. The caveat is that $k$ cannot be arbitrarily large due to limited register capacity. We store matrices in lower-triangular format, use in-place matrix factorizations and solves, and simplify the conditioning formulas to reduce register requirements to a minimum. The forward pass shown in Algorithm~\ref{alg:graphgp} requires only $(k+1)[d+1+(k+2)/2]$ temporary floats per conditioning step. Our target application tolerates a smaller number of neighbors, but for applications that require $k\gtrsim64$, the JAX version should be used.

\begin{figure}
  \centering
  \includegraphics[trim=0.3cm 1.0cm 0.4cm 1.0cm, clip, width=\linewidth]{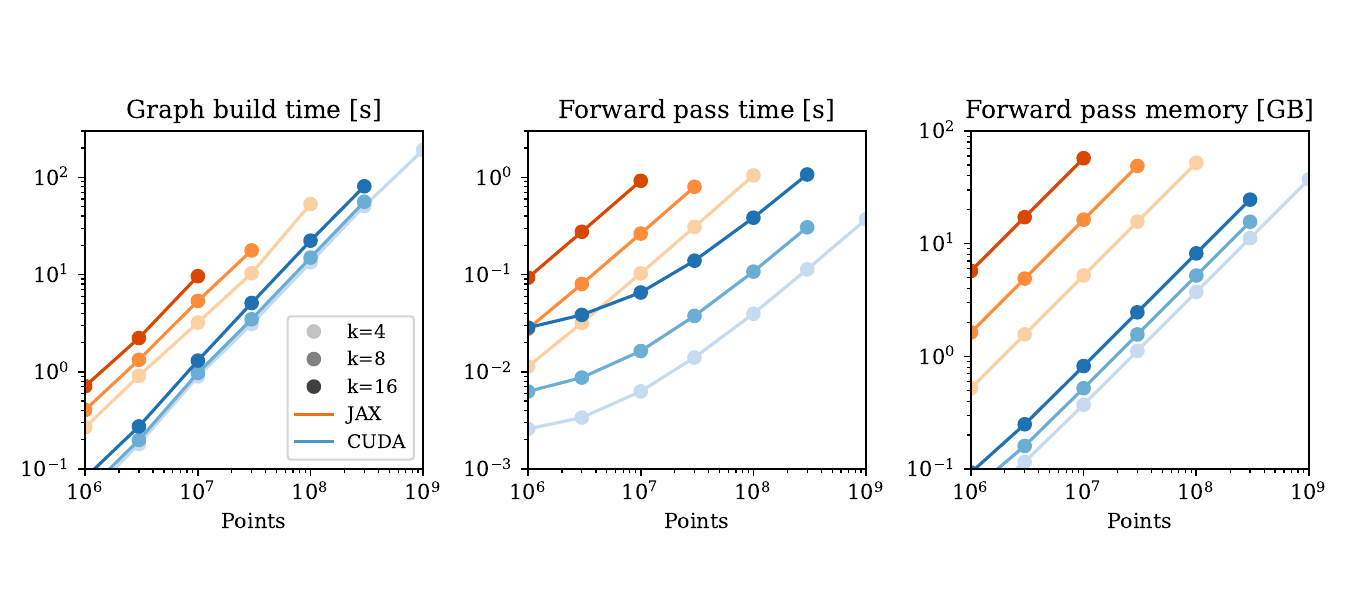}
  \caption{Performance of GraphGP as a function of $N$ and $k$. Points are drawn from a unit Gaussian in 3d with $n_0=1000$. All experiments were performed on an H100 GPU (80 GB) with 32-bit floats, increasing $N$ until memory was exhausted. The large improvements that CUDA offers for the forward pass are due to on-the-fly matrix operations described in Section~\ref{sec:implementation}.}
  \label{fig:speed-memory}
\end{figure}

Memory optimizations translate to performance improvements as our algorithm is memory-bound. Figure~\ref{fig:speed-memory} shows performance benchmarks for graph construction and the forward pass. Linear scaling with $N$ is achieved for the forward pass, and the CUDA implementation is more than an order of magnitude faster than the JAX implementation while requiring less memory.

Automatic differentiation is critical for fitting large models to data, which is often a reason to prefer JAX over custom high-performance code. Therefore, we implement forward and backward derivatives of GraphGP in CUDA as well. We use formulas for the derivatives of the Cholesky decomposition from \cite{smith1995} and register a custom JAX primitive so that GraphGP can be used seamlessly as a part of larger differentiable models. We test against the JAX implementation to verify correctness.

Graph construction requires building a $k$-d tree and querying neighbors, for which we use the parallel, nearly in-place, sort-based GPU algorithms from \cite{wald2022a, wald2022b}, modified for our custom order. Next, computing graph depths sequentially can be prohibitively slow, so we apply the depth update described in Section~\ref{sec:ordering} across all points in parallel and iterate until no depths change. While this requires more operations, it is almost always faster for our shallow graphs on modern GPUs.

\begin{algorithm}[t]
\caption{GraphGP forward pass}\label{alg:graphgp}
\begin{algorithmic}[1]
\Require Points $\mathbf{X} \in \mathbb{R}^{N \times d}$, covariance $\kappa(r)$, initial size $n_0$, neighbor indices $C \in \mathbb{N}^{(N-n_0)\times k}$, parallel batches $\mathcal{B}_1, \ldots, \mathcal{B}_L$, a partition of $\{n_0{+}1, \ldots, N\}$, white noise $\boldsymbol{\xi} \sim \mathcal{G}(0, I_N)$
\Ensure GP realization $\mathbf{v} \in \mathbb{R}^N$

\State $\mathbf{v}_{1:n_0} \gets \mathrm{chol}(\kappa(\|\mathbf{x}_{1:n_0} - \mathbf{x}_{1:n_0}^\top\|))\,\boldsymbol{\xi}_{1:n_0}$ \Comment{Dense initial batch}
\For{$\ell = 1, \ldots, L$}
  \ForAll{$i \in \mathcal{B}_\ell$ \textbf{in parallel}}
    \State $\mathbf{K} \gets \kappa(\|\mathbf{x}_{[c(i),\,i]} - \mathbf{x}_{[c(i),\,i]}^\top\|)$ \Comment{$(k{+}1)\times(k{+}1)$ joint covariance}
    \State $\mathbf{L} \gets \mathrm{chol}(\mathbf{K})$
    \State $v_i \gets \mathbf{L}_{k+1,\,1:k}\,\mathbf{L}_{1:k,\,1:k}^{-1}\,\mathbf{v}_{c(i)} + L_{k+1,\,k+1}\,\xi_i$ \Comment{Equivalent to Eq.~\ref{eq:conditional}}
  \EndFor
\EndFor
\end{algorithmic}
\end{algorithm}

\section{Discussion}
\label{sec:discussion}

GraphGP enables practical scaling of Vecchia's approximation to nearly a billion points on a single GPU. We achieve this through a point ordering that prioritizes scaling of neighbor searches and parallelization on accelerators, at minimal cost to approximation accuracy (Section~\ref{sec:ordering}). Attention to hardware capabilities in our CUDA implementation dramatically improves performance and reduces memory requirements, unlocking another order of magnitude in scale (Section~\ref{sec:implementation}). GraphGP thus opens the door to the use of Gaussian processes for three-dimensional mapping of the interstellar medium at high resolution and on arbitrary point distributions.

Related work includes \texttt{GpGpU} \citep{james2024}, which demonstrates a GPU implementation of Vecchia's approximation for $N\sim 10^6$ points. Crucially, \texttt{GpGpU} only implements the easily parallelized inverse and log-determinant, not the forward pass, and thus can only be used in a limited inference context. Commonly used algorithms such as Metric Gaussian Variational Inference \citep{knollmuller2019} and Geometric Variational Inference \citep{frank2021} require the forward pass for posterior sampling. Additionally, \texttt{GpGpU} does not provide a means to query neighbors on GPU, and the expensive CPU neighbor-finding procedure cannot scale to $N\sim 10^8$. GraphGP solves both problems using the custom order described in Section \ref{sec:ordering}, along with automatic differentiation and other performance improvements.

Iterative Charted Refinement \citep{edenhofer2022} uses a Vecchia-like approximation and scales to large $N$ by restricting points to a grid, from which neighbors are derived, and effectively severing dependencies in order to parallelize. GraphGP avoids these constraints with a $k$-d tree and graph analysis, and additionally provides stable kernel derivatives, an inverse, and a log-determinant. The result is the best of both Vecchia and ICR, enabling Gaussian processes to be used for a wider range of problems than was previously possible.


\bibliographystyle{unsrtnat}
\bibliography{references}

\end{document}